\documentclass[11pt]{article}
\title{Unsupervised Variable Selection for Ultrahigh-Dimensional Clustering Analysis}
\author{Tonglin Zhang\footnote{Department of Statistics, Purdue University, 150 North University Street, West Lafayette, IN 47907-2067, Email: tlzhang@purdue.edu} and Huyunting Huang\footnote{Department of Statistics, Purdue University, 150 North University Street, West Lafayette, IN 47907-2067, Email: huan1182@purdue.edu}}
\usepackage{mathrsfs}
\usepackage{amsfonts}
\usepackage{amsmath}
\usepackage{wrapfig,lipsum,booktabs}
\usepackage{bm}
\usepackage{algorithm}
\usepackage{algpseudocode}
\usepackage{pifont}
\usepackage{epsfig}
\def\qed{\hfill$\diamondsuit$}
\newtheorem{defn}{Definition}

\newtheorem{thm}{Theorem}
\newtheorem{cor}{Corollary}
\newtheorem{lem}{Lemma}
\begin{document}
\def\qed{\hfill$\diamondsuit$}
\def\ed{\end{document}}
\def\SSB{S\negthinspace S\negthinspace B}
\def\MSB{M\negthinspace S\negthinspace B}
\def\SSW{S\negthinspace S\negthinspace W}
\def\MSW{M\negthinspace S\negthinspace W}
\maketitle
\def\eqalign#1{\null\,\vcenter{\openup\jot\ialign
              {\strut\hfil$\displaystyle{##}$&$\displaystyle{{}##}$
               \hfil\crcr#1\crcr}}\,}

\begin{abstract}

Compared to supervised variable selection, the research on unsupervised variable selection is far behind. A forward partial-variable clustering full-variable loss (FPCFL) method is proposed for the corresponding challenges. An advantage is that the FPCFL method can distinguish active, redundant, and uninformative variables, which the previous methods cannot achieve. Theoretical and simulation studies show that the performance of a clustering method using all the variables can be worse if many uninformative variables are involved. Better results are expected if the uninformative variables are excluded. The research addresses a previous concern about how variable selection affects the performance of clustering. Rather than many previous methods attempting to select all the relevant variables, the proposed method selects a subset that can induce an equally good result. This phenomenon does not appear in the supervised variable selection problems.

\end{abstract}

{\it Key Words:}  Active Variables; Adjusted Rand Index;  Gaussian Mixture Models; Partial-variable Clustering Full-variable Loss; Full-variable Penalized Loss; Relevant, Redundant, and Uninformative Variables. 

AMS Classification: 62H30; 62J07; 62H35.

\section{Introduction}
\label{sec:introduction}

Recent advances in computer technologies bring challenges in ultrahigh-dimensional data analysis. Variable or feature selection is among the most important tools for the corresponding challenges. Compared to supervised variable selection, the research on unsupervised variable selection is far behind, leading to difficulty in ultrahigh-dimensional clustering analysis in practice. The proposed work devises a formal method to overcome the difficulty.

In the literature, variable selections are mostly studied under the framework of supervised machine learning (ML). A supervised ML problem assumes that the observed data contain a response variable and many explanatory variables. A supervised variable selection method is needed if the number of explanatory variables is large. In recent decades, the penalized maximum likelihood (PML) approach has been used to devise supervised variable selection methods. Examples include the LASSO~\cite{tibshirani1996}, the SCAD~\cite{fanli2001}, the MCP~\cite{zhang2010}, and the GIC~\cite{zhangli2010}. These methods cannot be applied to unsupervised ML problems due to the absence of the response. 

Clustering is one of the most important tasks in unsupervised ML. It partitions the data into many subsets, called clusters, such that observations within the clusters are the most homogeneous and observations between the clusters are the most heterogeneous. Examples of clustering methods include the $k$-means~\cite{macqueen1967,steinhaus1957}, the $k$-medians~\cite{cardot2012}, the $k$-modes~\cite{chaturvedi2001},  the generalized $k$-means~\cite{bock2008,zhanglin2021}, and the EM-algorithm~\cite{fraley2002,lau2007,nityasuddhi2003}. The $k$-means and the EM algorithm are the most straightforward. They belong to partitioning clustering, a category that can be interpreted by the centroidal Voronoi tessellation in mathematics~\cite{du2002}. Other categories include hierarchical clustering~\cite{zhao2005}, fuzzy clustering~\cite{trauwaert1991}, and density-based clustering~\cite{ester1996,kriegel2011}. Many of those have been incorporated into software packages, such as \textsf{R} and \textsf{Python}.

In ultrahigh-dimensional data analysis, clusters can often be determined by only a few variables. As noted by~\cite{milligan1989}, including uninformative variables could complicate clustering. The two broad approaches for identifying uninformative variables are variable weighting and variable selection~\cite{brusco2001}. Variable weighting uses all the variables. It assigns weights to variables for the task. Variable selection uses a subset of variables. It can be treated as a special case of variable weighting by assigning weights $0$ or $1$. Variable selection has obvious advantages over variable weighting because it eliminates unnecessary variables~\cite{fowlkes1988,gnanadesikan1995}.

Many unsupervised variable selection methods have been proposed and validated. In the comparison of eight methods, including the variable selection in $k$-means (VS-KM)~\cite{brusco2001}, the model-based variable selection~\cite{raftery2006}, the clustering objects on subsets of attributes (COSA)~\cite{friedman2004}, and the relative clusterability weighing~\cite{steinley2008a}, computational prohibition has been identified in the case when the number of variables is moderate due to exponential growth of the computational burden with the number of variables~\cite{steinley2008b}. This issue is overcome by several methods, such as the sparse $k$-means (SKM)~\cite{witten2010}, the sparse alternate sum (SAS)~\cite{arias-castro2017}, and the cardinality $k$-means (CKM)~\cite{yuan2022}. Other methods have also been proposed. Examples include the attribute weighting clustering (AWK)~\cite{chan2004}, the weighted $k$-means (WKM)~\cite{huang2005}, the entropy-weighting $k$-means (EW-KM)~\cite{jing2007}, the weighted locally clustering (WLC)~\cite{parvin2015}, and the rival penalized expectation-maximization (RPEM)~\cite{cheung2006}; to name a few. However, a recent study shows that most have not been comprehensively studied. There is a theoretical concern about how variable selection affects the performance of clustering~\cite{hancer2020}. We address this issue in our work.

The research has three main contributions. The first is the theoretical evaluation of unsupervised variable selection on the clustering performance. It addresses the previous concern about how variable selection affects clustering. The second is developing a forward partial-variable clustering full-variable loss (FPCFL) method to distinguish relevant, redundant, and uninformative variables. The third is non-uniqueness. We point out that in many cases, using a subset of relevant variables, called active variables, is enough for data clustering. The remaining are treated as redundant variables, implying that the set of active variables may not be uniquely determined.

The article is organized as follows. In Section~\ref{sec:background}, we review the background issues. In Section~\ref{sec:methodology}, we introduce our method. In Section~\ref{sec:simulation}, we evaluate the performance of our method with the comparison to our competitors by simulations. In Section~\ref{sec:application}, we apply our method to a real-world dataset. In Section~\ref{sec:discussion}, we provide a discussion. We put all of the proofs in the Appendix.

\section{Background}
\label{sec:background}

Terminologies are still inconsistent in unsupervised variable selection problems. We adopt the three terms called relevant, redundant, and uninformative variables used by~\cite{fop2018}. A relevant variable contains the essential information. Given a subset of selected relevant variables, another relevant variable is said redundant if it does not provide additional information for data clustering. An uninformative variable is irrelevant. It does not contain useful information.

We partition the relevant variables into a subset of active variables and a subset of redundant variables. Given the subset of the active variables, redundant variables can be ignored in clustering data. Our research aims to derive active, redundant, and uninformative variables. For instance, when only three variables are involved in the clustering problem reflected by Figure~\ref{fig:relevant, redundant, and uninformative variables}, both $x_1$ and $x_2$ are relevant. If $x_1$ is chosen as an active variable, then $x_2$ is a redundant variable; or vice versa. $x_3$ is an uninformative variable because the distribution is homogeneous between the clusters. The ideal case is to select $\{x_1\}$ or $\{x_2\}$ but not $\{x_1,x_2\}$. 

\begin{figure}
\centerline{\rotatebox{270}{\psfig{figure=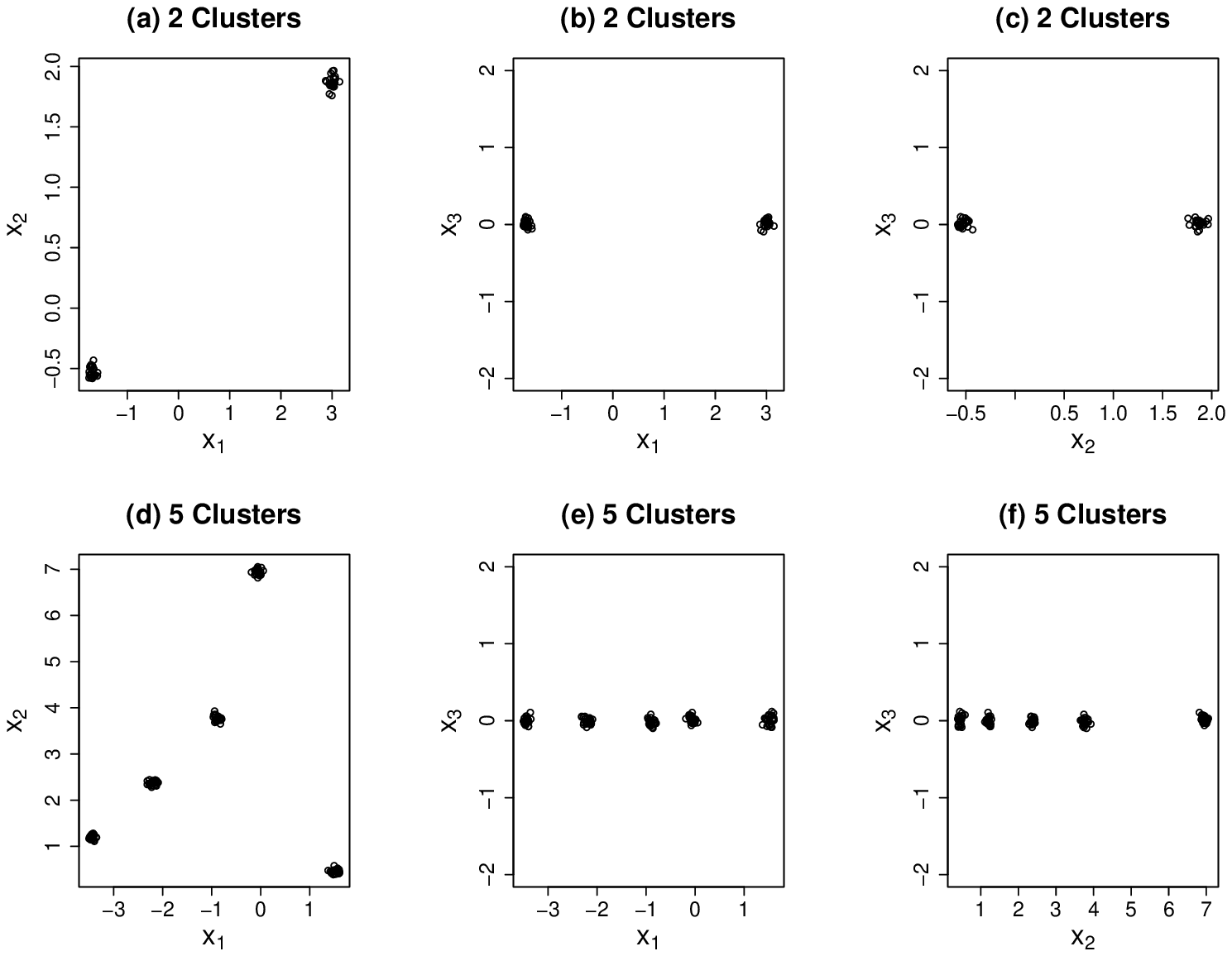,width=3.0in,}}}
\caption{\label{fig:relevant, redundant, and uninformative variables}Both $x_1$ and $x_2$ are relevant variables; $x_2$ is a redundant variable if $x_1$ is treated as an active variable (or vice versa); $x_3$ is an uninformative variable because its distribution is homogeneous between clusters.}
\end{figure}

Other terminology systems have also been proposed. An example is the terms relevant, irrelevant, and redundant variables used by~\cite{hancer2020,li2017}. They stand for the relevant, redundant, and uninformative variables displayed by Figure~\ref{fig:relevant, redundant, and uninformative variables}, respectively. 

The three tasks reflected by Figure~\ref{fig:relevant, redundant, and uninformative variables} include: (a) a subset of active variables that is adequate for data clustering; (b) a subset of redundant variables that can be ignored; and (c) a subset of the uninformative variables which can be removed. The previous methods cannot accomplish the three tasks simultaneously. For instance, the sparse clustering proposed by~\cite{witten2010} is unlikely to distinguish active and redundant variables. Similar issues also appear in a model-based variable selection method first proposed by~\cite{raftery2006} and later extended by~\cite{maugis2009a,maugis2009b} and another model-based variable selection method proposed by~\cite{marbac2017}. These methods often provide too many variables when $p$ is large. Because the computation of the global solution of typical clustering methods, such as the $k$-means and the EM-algorithm, is NP-hard~\cite{mahajan2012}, computation usually focuses on the local solutions, implying that their performance may be worse. A more convenient approach is to exclude the redundant and uninformative variables, such that a global solution is likely to be obtained. 

\section{Methodology}
\label{sec:methodology}

We review the well-known EM-algorithm and the $k$-means in Section~\ref{subsec:mixture models} with a motivated example in Section~\ref{subsec:motivated example}. We introduce our proposed method for selecting the active variables in Section~\ref{subsec:active variables} and identifying the redundant and uninformative variables in Section~\ref{subsec:redundant and uninformative variables}. We evaluate theoretical issues in Section~\ref{subsec:theoretical evaluation}.

\subsection{Gaussian Mixture Models and $k$-means}
\label{subsec:mixture models}

The EM-algorithm and the $k$-means are proposed under the Gaussian mixture models (GMMs). A GMM is a special case of mixture models with the underlying distribution to be the normal. Suppose that $n$ observations with $p$ variables are collected with data expressed as ${\mathcal D}=\{{\bm x}_i: i\in{\cal S}\}$, ${\mathcal S}=\{1,\dots,n\}$ and ${\bm x}_i=(x_{i1},\dots,x_{ip})^\top\in\mathbb{R}^p$. The size-$k$ clustering problem seeks to partition ${\cal S}$ into $k$ clusters as $\mathscr{C}_k=\{C_1,\dots,C_k: \bigcup_{s=1}^k C_s={\cal S}, C_s\not=\emptyset, C_{s}\cap C_{s'}=\emptyset, s\not=s' \}$, such that data within $C_s$ are the most homogeneous and data between distinct $C_{s}$ and $C_{s'}$ are the most heterogeneous. The EM-algorithm and the $k$-means assume that $k$ is usually unknown. A common way to determine the optimal $k$ is the Gap Statistics~\cite{tibshirani2001}.

A mixture model with $k$ components assumes that data are collected from $k$ sources with the distribution of the $s$th source as $f_s(\cdot)=f(\cdot;{{\bm\theta}_s})$, $s=1,\dots,k$, where ${\bm\theta}_s$ represents the parameters to be estimated from the data. A GMM assumes $f(\cdot;{{\bm\theta}_s})=\varphi(\cdot;{\bm\mu}_s,{\bm\Sigma}_s)$ with ${\bm\theta}_s=({\bm\mu}_s,{\bm\Sigma}_s)$, where $\varphi(\cdot;{\bm\mu}_s,{\bm\Sigma}_s)$ is the PDF of ${\cal N}({\bm\mu}_s,{\bm\Sigma}_s)$ with ${\bm\mu}_s=(\mu_{s1},\dots,\mu_{sp})^\top\in\mathbb{R}^{p}$ for the mean vector and ${\bm\Sigma}_s=(\sigma_{sj_1j_2})_{j_1,j_2=1,\dots,p}\in\mathbb{R}^{p\times p}$ for the covariance matrix. The three often used variants are the spherical GMM by ${\bm\Sigma}_s=\sigma^2{\bf I}$, linear discriminant analysis (LDA) GMM by ${\bm\Sigma}_s={\bm\Sigma}$, and quadratic discriminant analysis (QDA) GMM by ${\bm\Sigma}_s$ all distinct, respectively.

Let ${\bm z}_i=(z_{i1},\dots,z_{ik})^\top$ with $z_{is}\in\{0,1\}$ and $\sum_{s=1}^k z_{is}=1$ be the latent vector for the ground truths of ${\bm x}_i$, where $z_{is}=1$ if $i\in C_s$ or $z_{is}=0$ otherwise. The GMM assumes 
\begin{equation}
\label{eq:underlying distribution of the data incomplete data}
{\bm x}_i|{\bm z}_i\sim^{iid} \sum_{s=1}^k z_{is}\varphi({\bm x}_i;{{\bm\mu}_s},{\bm\Sigma}_s), {\bm z}_i\sim^{iid}\pi({\bm z}), 
\end{equation}
where $\pi({\bm z})$ is the prior distribution for the latent variable ${\bm z}_i$ specified by the Dirichlet distribution with the probability vector ${\bm\omega}=(\omega_1,\dots,\omega_k)^\top$. The marginal distribution of ${\mathcal D}$ satisfies
\begin{equation}
\label{eq:underlying distribution of the data complete data}
 {\bm x}_i\sim^{iid} \sum_{s=1}^k \omega_s \varphi({\bm x}_i;{{\bm\mu}_s},{\bm\Sigma}_s).
\end{equation}
The posterior distribution of the $j$th component of ${\bm z}_i$ given ${\mathcal D}$ is
\begin{equation}
\label{eq:posterior distribution of latent}
{\bm\pi}({\bm z}_i|{\bm x}_i)={ \pi({\bm z}_i) \sum_{s=1}^k z_{is}\varphi({\bm x}_i;{{\bm\mu}_s},{\bm\Sigma}_s)\over \sum_{s=1}^k \omega_s \varphi({\bm x}_i;{{\bm\mu}_s},{\bm\Sigma}_s)}.
\end{equation}

Because ${\bm\pi}({\bm z}_i|{\bm x}_i)$ cannot be solved analytically, numerical computation is used, leading to usable clustering methods in practice. A usable clustering method can only be developed under~\eqref{eq:underlying distribution of the data complete data}. The conditional distribution given by~\eqref{eq:underlying distribution of the data incomplete data} can only be used for theoretical evaluations. The EM-algorithm~\cite{dempster1977} is a method for fitting the GMM. The goal is the computing ${\bm\pi}({\bm z}_i|{\bm x}_i)$ given by~\eqref{eq:posterior distribution of latent}. Below, we briefly review the main steps of the two methods.

Let $\varphi(\cdot;{\bm\mu}_s^{(t-1)};{\bm\Sigma}_s^{(t-1)})$ and $\omega_s^{(t-1)}$ be the iterated values of $\varphi(\cdot;{\bm\mu}_s,{\bm\Sigma}_s)$ and $\omega_s$ provided by the previous (i.e., the $(t-1)$th) iteration for $s=1,\dots,k$, respectively. The E-step of the current (i.e., the $t$th) iteration predicts the $s$th component of ${\bm z}_i$ as 
\begin{equation}
\label{eq:E-step of EM}
 z_{is}^{(t)}={\omega_s^{(t-1)}\varphi({\bm x}_i;{\bm\mu}_s^{(t-1)},{\bm\Sigma}_s^{(t-1)})\over \sum_{j=1}^k \omega_j^{(t-1)}\varphi({\bm x}_i;{\bm\mu}_j^{(t-1)},{\bm\Sigma}_j^{(t-1)})} , s=1,\dots,k.
\end{equation}
The M-step updates the $s$ component of ${\bm\omega}$ by $\omega_s^{(t)}=(1/n) \sum_{i=1}^n z_{is}^{(t)}$ and ${\bm\theta}_s^{(t)}$. Applying the maximum likelihood approach conditioning on ${\bm\omega}^{(s)}=(\omega_1^{(t)},\dots,\omega_k^{(t)})^\top$, it has
\begin{equation}
\label{eq:estimate of the mean vector EM}
{\bm\mu}_s^{(t)}={\sum_{i=1}^n z_{is}^{(t)}{\bm x}_i\over\sum_{i=1}^n z_{ir}^{(t)}}
\end{equation} 
to update ${\bm\mu}_s$ in all the spherical, LDA, and QDA GMMs. In the QDA GMM, the M-step updates ${\bm\Sigma}_s$ by ${\bm\Sigma}_s^{(t)}=\sum_{i=1}^n z_{is}^{(t)}({\bm x}_i-{\bm\mu}_s^{(t)})^\top({\bm x}_i-{\bm\mu}_s^{(t)})/\sum_{i=1}^n z_{is}^{(t)}$ for $s=1,\dots,k$, respectively. In the LDA GMM, the M-step updates ${\bm\Sigma}$ by ${\bm\Sigma}^{(t)}=\sum_{s=1}^k\sum_{i=1}^n z_{is}^{(t)}({\bm x}_i-{\bm\mu}_s^{(t)})^\top({\bm x}_i-{\bm\mu}_s^{(t)})/n$. In the spherical GMM, the M-step updates $\sigma^2$ by $\sigma^{(t)2}= \sum_{s=1}^k\sum_{i=1}^n z_{is}^{(t)}\|{\bm x}_i-{\bm\mu}_s^{(t)}\|^2/(np)$. In the end, the EM-algorithm estimates the partition by
\begin{equation}
\label{eq:the partition EM}
\hat C_{s}=\{i: \hat z_{ir}=\mathop{\arg\!\max}_{j\in\{1,\dots,k\}} \hat z_{ij}\},\ s=1,\dots,k,
\end{equation}
where $\hat{\bm z}_i=(\hat z_{i1},\dots,\hat z_{ik})^\top$ is the imputation of ${\bm z}_i$. It treats~\eqref{eq:the partition EM} as an estimator of $\pi({\bm z}_i|{\bm x}_i)$ given by~\eqref{eq:posterior distribution of latent}. The formula is applied to every $i\in{\mathcal S}$ for data clustering. 

The $k$-means focuses on the spherical GMM. It uses
\begin{equation}
\label{eq:k-means criterion}
\widehat{\mathscr{C}_k}=\{\hat C_1,\dots,\hat C_k\}=\mathop{\arg\!\min}_{\mathscr{C}_k}Q(\mathscr{C}_k),
\end{equation}
where $Q({\mathscr{C}}_k)=\sum_{s=1}^k\sum_{i\in{ C_s}}\|{\bm x}_i-{\bm\mu}_s\|^2$ is the sum-of-squares (SSQ) criterion and ${\bm\mu}_s\in\mathbb{R}^p$ is the centroid of $C_s$, to compute $\pi({\bm z}_i|{\bm x}_i)$ for each $i\in{\mathcal S}$. Based on the previous centroids ${\bm\mu}_1^{(t-1)},\dots,{\bm\mu}_k^{(t-1)}$, the $k$-means imputes $s$th component of ${\bm z}_i$ in the current iteration as
\begin{equation}
\label{eq:prediction of ground truth latent variables}
z_{is}^{(t)}=\left\{ \begin{array}{cc}   1,  &  s=\mathop{\arg\!\min}_{j\in\{1,\dots,k\}}\|{\bm x}_i-{\bm\mu}_j^{(t-1)}\| , \cr 0 , & {\rm otherwise}, \end{array}\right. 
\end{equation}
and the $s$th cluster in the current iteration as
\begin{equation}
\label{eq:the rth cluster in the iterations}
C_s^{(t)}=\{i: z_{is}^{(t)}=1\}, s=1,\dots,k.
\end{equation}
It then updates the current centroids as
\begin{equation}
\label{eq:estimator of mu k-means}
{\bm\mu}_s^{(t)}={1\over |C_s^{(r)}|}\sum_{i\in C_s^{(t)}} {\bm x}_i, s=1,\dots,k.
\end{equation}
The $k$-means uses~\eqref{eq:prediction of ground truth latent variables},~\eqref{eq:the rth cluster in the iterations}, and~\eqref{eq:estimator of mu k-means} to compute the estimator of $\pi({\bm z}_i|{\bm x}_i)$ for every $i\in{\mathcal S}$. 

\subsection{Motivated Example}
\label{subsec:motivated example}

We present a motivated example under the $k$-means. We consider the simplified case which has ${\bm z}_i^{(0)}={\bm z}_i$ at the beginning. We treat it as an Oracle initialization because it assumes that an Oracle knows the true labels and provides them for the initialization. We show that the partition given by the first iteration tends to be random when many uninformative variables are used.

We assume that only a small portion of ${\bm\mu}_s=(\mu_{s1},\dots,\mu_{sp})^\top$ are distinct such that they satisfy
\begin{equation}
\label{eq:assumption of spatsity}
\exists q\ll p\ s.t.\ \mu_{1j}=\cdots=\mu_{kj}\ \forall j>q,
\end{equation}
implying that ${\bm\mu}_s$, $s=1,\dots,k$, can only be different in their first $q$ components. We denote that the true cluster label of ${\bm x}_i$ as $s_i$, implying $z_{is_i}=1$ and $z_{is_i'}=0$ for any $s_i'\not=s_i$. Clustering assignment of the $k$-means for ${\bm x}_i$ in the next iteration is determined by $Q_i=\|{\bm x}_{i}-\hat{\bm\mu}_{s_i}\|-\|{\bm x}_{i}-\hat{\bm\mu}_{s_i'}\|^2$, where $\hat{\bm\mu}_{s_i}=\sum_{i=1}^n z_{is_i}{\bm x}_i/n_{s_i}$ is the estimated centroid of cluster $s_i$ and $n_{s_i}=\sum_{i=1}^n z_{is_i}$ is the size of cluster $s_i$. By ${\bm x}_{i}-\hat{\bm\mu}_{s_i}\sim {\cal N}[{\bm 0},\sigma^2(1-1/n_{s_i}){\bf I}]$, ${\bm x}_{i}-\hat{\bm\mu}_{s_i'}\sim {\cal N}[{\bm\mu}_{s_i}-{\bm\mu}_{s_i'}, \sigma^2(1+1/n_{s'}) {\bf I}]$, and ${\rm cov}({\bm x}_{i}-\hat{\bm\mu}_{s_i},{\bm x}_{i}-\hat{\bm\mu}_{s_i'})=\sigma^2(1-1/n_{s_i}){\bf I}$, we obtain
\begin{equation}
\label{eq:expected value for next assignment k-means}
{\rm E}(Q_i)=-p\sigma^2 (1/n_{s_i}+1/n_{s_i'})-\sum_{j=1}^q (\mu_{s_ij}-\mu_{s_i'j})^2
\end{equation}
and
\begin{equation}
\label{eq:variance for next assignment k-means}
{\rm var}(Q_i)=2p\sigma^4[3(1-1/n_{s_i})^2+(1+1/n_{s_i'})^2]+4\sigma^2\sum_{j=1}^q (\mu_{s_ij}-\mu_{s_i'j})^2.
\end{equation}

If $p$ is large, then the second terms on the right-hand sides of~\eqref{eq:expected value for next assignment k-means} and~\eqref{eq:variance for next assignment k-means} can be ignored, leading to ${\rm E}(Q_i)\approx -p\sigma^2 (1/n_{s_i}+1/n_{s_i'})$ and ${\rm var}(Q_i)\approx2\sigma^4[3p(1-1/n_{s_i})^2+p(1+1/n_{s_i'})^2]$, which do not depend on ${\bm\mu}_{s_i}$ and ${\bm \mu}_{s_i'}$. The differences between ${\bm\mu}_s$ for $s=1,\dots,k$ vanish. Clustering assignments tend to be random if all the variables are used. If the first $q$ variables can be identified by an unsupervised variable selection method, then the first terms on the right-hand sides of~\eqref{eq:expected value for next assignment k-means} and~\eqref{eq:variance for next assignment k-means} are reduced. The contribution of the disparity between ${\bm\mu}_s$, $s=1,\dots,k$, is amplified. The $k$-means performs worse if the uninformative variables are used. It is critical to exclude all the uninformative variables from the $k$-means. This property is strictly proven by Theorem~\ref{thm:randomly assignments} of Section~\ref{subsec:theoretical evaluation}. It is also illustrated. 

\subsection{Active Variables}
\label{subsec:active variables}

The proposed method selects the active, redundant, and uninformative variables from the variable set denoted as ${\mathcal A}=\{1,\dots,p\}$, respectively. It the uses the active variables to cluster the data. We need official definitions of the active, redundant, and uninformative variables. They are defined based on  $\pi({\bm z}_i|{\bm x}_i)$ given by~\eqref{eq:posterior distribution of latent}.

\begin{defn}
\label{defn:relevant, redundant, and uninformative}
 Let ${\mathcal A}_r$ and ${\mathcal A}_u$ be a partition of ${\mathcal A}$ such that it induces ${\bm x}_i=({\bm x}_{ir}^\top,{\bm x}_{iu}^\top)^\top$ up to a permutation of the lags in ${\mathcal A}$, where ${\bm x}_{ir}\in\mathbb{R}^{p_r}$ and ${\bm x}_{iu}\in\mathbb{R}^{p_u}$ with $p_r=|{\mathcal A}_r|$ and $p_u=|{\mathcal A}_u|$ satisfying $p_r+p_u=p$. We say ${\mathcal A}_u$ is uninformative and ${\mathcal A}_r$ is relevant if $\pi({\bm z}_i|{\bm x}_i)=\pi({\bm z}_i|{\bm x}_{ir})$ and $\pi({\bm z}_i|{\bm x}_{iu})=\pi({\bm z}_i)$. We say ${\mathcal A}_u$ is maximum uninformative if it is uninformative but ${\mathcal A}_u\cup\{j\}$ is not for any $j\in{\mathcal A}_r$. If ${\mathcal A}_u$ is maximum uninformative, then the corresponding ${\mathcal A}_r$ is said minimum relevant. 
\end{defn}

The determination of ${\mathcal A}_r$ and ${\mathcal A}_u$ does not depend on $i$ because ${\bm z}_i|{\bm x}_i$ are iid. The distribution of ${\bm z}_i|{\bm x}_{ir}$ as $\pi({\bm z}_i|{\bm x}_{ir})$ is derived by integrating out ${\bm x}_{iu}$ from the distribution of $({\bm z}_i^\top,{\bm x}_{iu}^\top)^\top|{\bm x}_{ir}$. The distribution of ${\bm z}_i|{\bm x}_{iu}$ is derived by integrating our ${\bm x}_{ir}$ from the distribution of $({\bm z}_i^\top,{\bm x}_{ir}^\top)^\top|{\bm x}_{iu}$. Both are well-defined under~\eqref{eq:underlying distribution of the data incomplete data}. An uninformative ${\mathcal A}_u$ needs to satisfy both ${\bm z}_i\perp {\bm x}_{iu}|{\bm x}_{ir}$ and ${\bm z}_i\perp {\bm x}_{iu}$ simultaneously. It is still uninformative if a variable from ${\mathcal A}_u$ is removed but may not if a variable in ${\mathcal A}_r$ is added. Therefore, we need the definition of the maximum uninformative set.

\begin{thm}
\label{thm:uninformative GMM}
${\mathcal A}_u$ is uninformative in a GMM iff ${\rm E}({\bm x}_{iu}|{\bm z}_i)$ and ${\rm cov}({\bm x}_{iu}|{\bm z}_i)$ do not depend on ${\bm z}_i$. The maximum uninformative set of a GMM is unique.
\end{thm}

The conclusion of Theorem~\ref{thm:uninformative GMM} cannot be extended to arbitrary mixture models due to the issue that pairwise and mutual independence are not equivalent if the underlying distribution is not normal. Thus, we restrict our attention to the GMMs in our methodological development. 

All the uninformative variables should be excluded from the analysis because they only increase randomness in data clustering. It is important to restrict the clustering analysis on the relevant variables to enhance the accuracy. In many cases, it is not necessary to use all the relevant variables because the usage of a subset of those can be equally better. We then define the active and redundant variables. 

\begin{defn}
\label{defn:definition of redundent variable}
Let ${\mathcal A}_r$ be the minimum relevant variable set and ${\bm x}_{ir}$ be the sub-vector of ${\bm x}_i$ composed of variables in ${\mathcal A}_r$. Suppose that up to a permutation of the lags there is ${\bm x}_{ir}=({\bm x}_{ia}^\top,{\bm x}_{ir_{ed}}^\top)^\top$, where ${\bm x}_{ia}$ and ${\bm x}_{ir_{ed}}$ are sub-vectors of ${\bm x}_{ir}$ composed of variables contained in a partition ${\mathcal A}_a$ and ${\mathcal A}_{r_{ed}}$ of ${\mathcal A}_{r}$, respectively. We say ${\mathcal A}_{r_{ed}}$ is redundant and ${\mathcal A}_a$ is active if ${\bm z}_i\perp{\bm x}_{ir_{ed}}|{\bm x}_{ia}$. We say ${\mathcal A}_{r_{ed}}$ is maximum redundant if it is redundant but ${\mathcal A}_{r_{ed}}\cup\{j\}$ is not for any $j\in {\mathcal A}_a$. If ${\mathcal A}_{r_{ed}}$ is maximum redundant, then the corresponding ${\mathcal A}_a$ is said minimum active. 
\end{defn}

Similar to Definition~\ref{defn:relevant, redundant, and uninformative}, the determination of ${\mathcal A}_a$ and ${\mathcal A}_{r_{ed}}$ does not depend on $i$ either. The choice of the redundant and active variable sets is not unique. We may have that ${\mathcal A}_{r_2}$ is redundant when ${\mathcal A}_{r_1}$ is treated as active or vice versa for a partition ${\mathcal A}_{r_1}$ and ${\mathcal A}_{r_2}$ of ${\mathcal A}_r$. If ${\mathcal A}_{a}$ is a minimum active variable set, we may derive another minimum active variable set by $({\mathcal A}_{a}\backslash\{j\})\cup\{j'\}$ for some $j\in{\mathcal A}_{a}$ and $j'\in{\mathcal A}_{r}\backslash{\mathcal A}_{a}$. An example is the case displayed in Figure~\ref{fig:relevant, redundant, and uninformative variables}. The minimum relevant variable set is ${\mathcal A}_{r}=\{1,2\}$. We can treat ${\mathcal A}_a=\{1\}$ as a minimum active variable set if $\{2\}$ is treated as a redundant variable set. We derive another minimum active variable set by $\{2\}=({\mathcal A}_a\backslash\{1\})\cup\{2\}$ with $j=1$ and $j'=2$.

We propose an unsupervised variable selection method to select active, redundant, and uninformative variables from ${\mathcal A}$. We call it an FPCFL method because it is a forward unsupervised variable selection method with a partial-variable clustering (PC) stage to partition the data and a full-variable loss (FL) stage to measure the loss. The PC stage uses the partial variable set of data denoted as ${\mathcal D}_\alpha=\{{\bm x}_{i\alpha}:i\in{\mathcal S}\}$ to cluster data, where ${\bm x}_{i\alpha}$ is the sub-vector of ${\bm x}_i$ composed of variables contained in a candidate $\alpha\subset{\mathcal A}$ to be considered as an active variable set in our method. Suppose that a partition of ${\mathcal S}$ denoted as $\hat{\mathscr C}_{k\alpha}=\{\hat C_{1\alpha},\dots,\hat C_{k\alpha}\}$ is provided by a clustering method based on ${\mathcal D}_\alpha$. The FL stage measures the loss of $\hat{\mathscr C}_{k\alpha}$ based on the full variable set (i.e., the entire ${\mathcal D})$). We devise three kinds of loss functions for the FL stage according to the three variants of the GMMs, respectively. 

If the $k$-means or the EM-algorithm for the spherical GMM is chosen as the clustering method, then the FL stage uses the full-variable penalized  loss (FPL) as
\begin{equation}
\label{eq:penalized likelihood loss spherical}
{\mathcal L}_\lambda(\hat{\mathscr C}_{k\alpha})=-2\ell(\hat{\mathscr C}_{k\alpha})+\lambda df_\alpha,
\end{equation}
where $\ell(\hat{\mathscr C}_{k\alpha})=-(np/2)[1+\log(2\pi)]-(np/2)\log(\hat\sigma_\alpha^2)$ is the loglikelihood function of the spherical GMM under the partition given by $\hat{\mathscr C}_{k\alpha}$, $\hat\sigma_\alpha^2=(np)^{-1}\sum_{s=1}^k\sum_{i\in\hat C_{s\alpha}}\|{\bm x}_i-\hat{\bm\mu}_{s\alpha}\|^2$ is the corresponding estimator of $\sigma^2$, $\hat{\bm\mu}_{s\alpha}=(1/|\hat C_{s\alpha}|)\sum_{i\in\hat C_{s\alpha}}{\bm x}_i$ is the corresponding estimator of ${\bm\mu}_s$, $\lambda>0$ is a tuning parameter, and $df_\alpha=k|\alpha|$ is the corresponding degrees of freedom. We show that it is inappropriate to choose $\lambda=0$ (i.e., Corollary~\ref{cor:compared with random clustering unsupervised variable selection}). Therefore, the tuning parameter must be positive in~\eqref{eq:penalized likelihood loss spherical}. 

If the EM-algorithm for the LDA GMM is chosen as the clustering method, then the FPL becomes
\begin{equation}
\label{eq:penalized likelihood loss lda}
{\mathcal L}_{\lambda,LDA}(\hat{\mathscr C}_{k\alpha})=-2\ell_{LDA}(\hat{\mathscr C}_{k\alpha})+\lambda df_{\alpha,LDA},
\end{equation}
where $\ell_{LDA}(\hat{\mathscr C}_{k\alpha})=-(np/2)[1+\log(2\pi)]-(k/2)\log(\hat{\bm\Sigma}_\alpha)$ is the loglikelihood function of the LDA GMM under the partition given by $\hat{\mathscr C}_{k\alpha}$, $\hat{\bm\Sigma}_\alpha=(np)^{-1}\sum_{s=1}^k\sum_{i\in\hat C_{s\alpha}}({\bm x}_i-\hat{\bm\mu}_{s\alpha})({\bm x}_i-\hat{\bm\mu}_{s\alpha})^\top$ is the corresponding estimator of ${\bm\Sigma}$, and $df_{\alpha,LDA}=k|\alpha|+|\alpha|(|\alpha|+1)/2$ is the corresponding degrees of freedom. The FPL of the EM-algorithm for the QDA GMM can also be derived. Because most of the previous methods are proposed for the spherical GMM, we focus on our presentation based on~\eqref{eq:penalized likelihood loss spherical}. 

The formulation of the FPL given by~\eqref{eq:penalized likelihood loss spherical} does not rely on the clustering method used in the PC stage. It can be applied to an arbitrary clustering method for partitioning ${\mathcal S}$ based on ${\mathcal D}_\alpha$ beyond the EM-algorithm and the $k$-means. The first term on the right-hand side of~\eqref{eq:penalized likelihood loss spherical} controls the partition loss. It does not change if the partition remains the same. The second term controls the number of variables. It becomes larger if the same partition is induced by using more variables. We select the variable set by
\begin{equation}
\label{eq:selection of variable set}
\hat\alpha=\mathop{\arg\!\min}_{\alpha} \ell_\lambda(\hat{\mathscr C}_{k\alpha})
\end{equation}
under the spherical GMM. Similar methods can be devised for the LDA and QDA GMMs. To apply~\eqref{eq:selection of variable set}, it is necessary to provide an appropriate $\lambda$. We adopt the well-known BIC approach, leading to $\lambda=\log(np)$. 

If the brute-force approach is used, there are as many as $2^p$ candidates of $\alpha$ to be considered in~\eqref{eq:selection of variable set}. This is treated as an impossible task if $p$ is not small. We propose a forward partial-variable clustering full-variable loss (FPCFL) algorithm to overcome the difficulty. The FPCFL algorithm reduces the computational complexity from $2^p$ to $p^2$. It can be used even if $p$ is large.

The FPCFL algorithm starts with the empty set and adds one variable each time until the FPL value cannot be reduced. Our method chooses $\alpha^{(0)}=\emptyset$ at the beginning, where $\alpha^{(t)}$ is the iterated $\hat\alpha$ given by the $t$th iteration. Based on $\alpha^{(t-1)}$ given by the $(t-1)$th iteration, the $t$th iteration calculates the most important variable by 
\begin{equation}
\label{eq:the most important feature first step}
j_{\min}^{(t)}=\mathop{\arg\!\min}_{j\not\in\alpha^{(t-1)}} {\cal L}_{\lambda}(\hat{\mathscr{C}}_{\alpha^{(t-1)}\cup\{j\}}).
\end{equation}
It updates $\alpha^{(t-1)}$ by $\alpha^{(t)}=\alpha^{(t-1)}\cup\{j_{\min}^{(t)}\}$ if $\ell(\hat{\mathscr{C}}_{\alpha^{(t)}})<\ell(\hat{\mathscr{C}}_{\alpha^{(t-1)}})$; or the algorithm stops otherwise. We obtain Algorithm~\ref{alg:variable selection for unsupervised machine learning}.

\begin{algorithm}
\caption{\label{alg:variable selection for unsupervised machine learning} The FPCFL algorithm for $\hat\alpha$ given by~\eqref{eq:selection of variable set}}
\begin{algorithmic}[1]
\Statex{{\bf Input}: Data set ${\cal D}=\{{\bm x}_1,\dots,{\bm x}_n\}$ and the number of clusters $k$}
\Statex{{\bf Output}: $\hat\alpha$ and the corresponding $\hat {\bm z}_1,\dots,\hat{\bm z}_n$}
\Statex{\it Initialization}
\State{Let $\alpha^{(0)}=\emptyset$}
\Statex{\it Begin Iteration}
\State{Let $\alpha^{(t-1)}=\{j_{\min}^{(1)},j_{\min}^{(2)},\dots,j_{\min}^{(t-1)}\}$ be given by the $(t-1)$th iteration}
\State{Compute the current $j_{\min}^{(t)}$ by~\eqref{eq:the most important feature first step}}
\State{$\alpha^{(t)}\leftarrow\alpha^{(t-1)}\cup \{j_{\min}^{(t)}\}$ if ${\cal L}_{\lambda}(\mathscr{C}_{\alpha^{(t-1)}\cup\{j_{\min}^{(t)}\}})<{\cal L}_{\lambda}(\mathscr{C}_{\alpha^{(t-1)}}) $ or stops otherwise}
\Statex{\it End Iteration}
\State {Output}
\end{algorithmic}
\end{algorithm}

The iterations of Algorithm~\ref{alg:variable selection for unsupervised machine learning} do not remove variables. If an iteration does not add a variable, then the algorithm stops. The stopping rule is determined by combining the two terms on the right-hand side of~\eqref{eq:penalized likelihood loss spherical}. If the reduction of the loglikelihood function specified by the first term on the right-hand side of~\eqref{eq:penalized likelihood loss spherical} can be ignored, then the second term dominates the changes, leading to an increase of the FPL with the output of $\hat\alpha$ to be the variable set given by the previous iteration; otherwise, the first term dominates the change and the variable is added, leading to the next iteration. Therefore, the minimum active variable set will likely be the output of Algorithm~\ref{alg:variable selection for unsupervised machine learning}.  

The choice of the number of clusters (i.e., $k$) is an important issue. It can be addressed by combining Algorithm~\ref{alg:variable selection for unsupervised machine learning} with a previous method for determining the number of clusters. An example is the Gap Statistics~\cite{tibshirani2001} that has been extensively used in determining the number of clusters in the literature. We incorporate the Gap Statistics into our method. The approach is straightforward. We calculate the Gap Statistics values based on each selected $k$. We determine the best $k$ by minimizing the Gap Statistics. Our method can be applied even if $k$ is unknown. 

We compare our method with the previous methods. One of the main differences we find is the evaluation of the loss. All the variables are used in measuring the loss in our method. Only the selected variables are used in the previous methods. We treat the previous as the partial-variable clustering partial-variable loss (PCPL) methods. For instance, the \textsf{clustvarsel} method proposed by~\cite{raftery2006} uses the BIC to measure the importance of a candidate variable set. The formulation of the BIC objective function does not contain any variables outside of the candidate variable set. A similar issue also occurs in the \textsf{VarSelLCM} method proposed by~\cite{marbac2017}, the \textsf{SalvarMix} method proposed by~\cite{maugis2009b}, and the \textsf{vscc} method proposed by the \textsf{vscc} package of \textsf{R}. An exception is the \textsf{sparcl} method proposed by~\cite{witten2010} that aims to provide weights for the importance of variables via a sparse clustering framework. The \textsf{sparcl} indirectly selects the variables by eliminating those with weights lower than a threshold value. The different strategy for evaluating the loss means that we focus on choosing the active variables while the previous methods focus on the union of the active and redundant variables. 

We illustrate the difference between our proposed and the previous methods by Figure~\ref{fig:relevant, redundant, and uninformative variables}. The results of $\hat{\mathscr C}_{k\alpha}$ are identical if we choose $\alpha$ as any nonempty subset of $\{1,2,3\}$ but not $\{3\}$. The second term on the right-hand of~\eqref{eq:penalized likelihood loss spherical} implies that we should use less variables, leading to $\hat\alpha=\{1\}$ or $\{2\}$ in our method. The previous \textsf{clustvarsel}, \textsf{VarSelLCM}, \textsf{SalvarMix}, and \textsf{vscc} methods likely provide $\hat\alpha=\{1,2\}$ as their answers because both the first and the second variables are important in data clustering. The \textsf{sparcl} method indirectly selects the variables by assigning large weights to the first and the second variables and a small weight to the third, which also likely provides $\{1,2\}$ as its final answer. We treat this property as a significant difference between our proposed and the previous methods. 

\subsection{Redundant and Uninformative Variables}
\label{subsec:redundant and uninformative variables}

Redundant or uninformative variables are unlikely to be contained in $\hat\alpha$ given by~\eqref{eq:selection of variable set}. By treating all variables contained in $\hat\alpha$ as active, we distinguish $j\not\in\hat\alpha$ as a redundant or uninformative variable by testing its relationship to the resulting partition $\hat{\mathscr C}_{k\hat\alpha}$ of ${\mathcal S}$ given by the clustering method based on the active variables claimed by $\hat\alpha$. For each $j\not\in\hat\alpha$, we assess
\begin{equation}
\label{eq:hypothesis testing the difference of the means}
H_0: \mu_{1j}=\cdots=\mu_{kj} \leftrightarrow H_1: \mu_{sj}\not=\mu_{s'j}~\exists~s,s'\in\{1,\dots,k\}
\end{equation}
by an F-test under the assumption that $x_{ij}|\hat z_{is}=1\sim{\mathcal N}(\mu_{sj},\sigma^2)$ independently, where $\hat{\bm z}_{i}=(\hat z_{i1},\dots,z_{ik})^\top$ is the cluster assignment vector claimed by $\hat{\mathscr C}_{k\hat\alpha}$. We use the traditional F-statistic for~\eqref{eq:hypothesis testing the difference of the means} as
\begin{equation}
\label{eq:the F-statistic for redundant and uninformative variables}
F_j^*={  \sum_{s=1}^k  (|\hat C_{s\alpha}|-1)(\hat\mu_{sj}- \hat\mu_{\cdot j})^2/(n-k)\over\sum_{s=1}^k\sum_{i\in\hat C_{s\alpha}}(x_{ij}-\hat\mu_{sj})^2/(|\hat C_{s\alpha}|-1) },
\end{equation}
where $\hat\mu_{sj}=\sum_{i\in\hat C_{s\alpha}} x_{ij}/|C_{s\alpha}|$ and $\hat\mu_{\cdot j}=\sum_{s=1}^k \sum_{i\in\hat C_{s\alpha}} x_{ij}/n$. By $F_j^*\sim F_{k-1,n-k}$ under $H_0$, we draw that variable $j$ is redundant if $F_j^*>c$ or uninformative otherwise, where $c$ is the critical value. We use the Bonferroni method to account for the multiple testing problem caused by the number of variables not contained in $\hat\alpha$, leading to $c=F_{0.05/(p-|\hat\alpha|),k-1,n-k}$ at $0.05$ significance level. We claim $j\not\in\hat\alpha$ is redundant if $F_j^*>F_{0.05/(p-|\hat\alpha|),k-1,n-k}$ or uninformative otherwise. Our method can also distinguish the redundant and uninformative variables. 

\subsection{Theoretical Evaluation}
\label{subsec:theoretical evaluation}

We investigate a few theoretical issues under the framework of GMMs. Although various kinds of GMMs can be considered (e.g., those in \textsf{mclust} package of \textsf{R}), we focus on the evaluations of the spherical, LDA, and QDA GMMs due to their popularity. We assume that~\eqref{eq:assumption of spatsity} is satisfied. We show that the partition given by a clustering method tends to be random under $p\gg q$ when all variables are used. We study the EM-algorithm and the $k$-means for the spherical GMM at the beginning. We then move our interest to the EM-algorithm for the LDA and QDA GMMs. 

 For the spherical GMM, the label assignments given by the $t$th iteration are determined by $\|{\bm x}_i-{\bm\mu}_s^{(t)}\|$ for $s=1,\dots,k$ and $i=1,\dots,n$. The label assignment for ${\bm x}_i$ is determined by
\begin{equation}
\label{eq:differece of the distance between two centers in the spherical model}
d_{ss'i}=\|{\bm r}_{si}\|^2-\|{\bm r}_{s'i}\|^2={\bm y}_{ss'i}^\top{\rm diag}({\bf I},-{\bf I}){\bm y}_{ss'i}
\end{equation}
for distinct $s,s'\in\{1,\dots,k\}$, where ${\bm y}_{ss'i}=({\bm r}_{si}^\top,{\bm r}_{s'i}^\top)^\top$, ${\bm r}_{si}={\bm x}_i-{\bm\mu}_s^{(t)}$, ${\bm r}_{s'i}={\bm x}_i-{\bm\mu}_s^{(t)}$, and ${\bm\mu}_s$ and ${\bm\mu}_{s'}$ are given by~\eqref{eq:estimate of the mean vector EM}. The label assignments ${\bm z}_1^{(t)},\dots,{\bm z}_n^{(t)}$ are given in the previous iteration. The M-step of the EM-algorithm treats those as fixed. It uses ${\bm z}_i^{(t)}$ for $i=1,\dots,n$ to compute ${\bm\mu}_s^{(t)}$ for $s=1,\dots,k$. The E-step uses ${\bm\mu}_s^{(t)}$ for $s=1,\dots,k$ to compute the new label assignments, leading to ${\bm z}_1^{(t+1)},\dots,{\bm z}_n^{(t+1)}$ for the next iteration. 

We evaluate the theoretical properties of ${\rm E}(d_{ss'i})$ and ${\rm var}(d_{ss'i})$ under the assumption that the true cluster assignment for ${\bm x}_i$ is $s$, leading to
${\bm x}_i\sim{\cal N}({\bm\mu}_s,\sigma^2{\bf I})$ independently. We assume that the cluster assignment given by the previous iteration is $s'$. Because $s'$ may be different from $s$, the true distribution of ${\bm x}_i-{\bm\mu}_s^{(t)}$ may be different from ${\cal N}({\bm\mu}_s,\sigma^2{\bf I})$. We compute ${\rm E}({\bm y}_{ss'i})$ and ${\rm cov}({\bm y}_{ss'i})$. We obtain ${\rm E}({\bm r}_{si})={\bm\nu}_{ssi}$ and ${\rm E}({\bm r}_{s'i})={\bm\nu}_{ss'i}$, where ${\bm\nu}_{ss'i}={\bm\mu}_{s}-{\sum_{j=1}^n z_{js'}^{(t)}{\bm\mu}_j/\sum_{j=1}^n} z_{js'}^{(t)}$. Moreover, we have ${\rm var}({\bm r}_{si})=a_{si}\sigma^2{\bf I}$, ${\rm var}({\bm r}_{s'i})=a_{s'i}\sigma^2{\bf I}$, and ${\rm cov}({\bm r}_{si}, {\bm r}_{s'i})=b_{ss'i}\sigma^2{\bf I}$, where 
\begin{equation}
\label{eq:definition of variance difference}
a_{si}=\left(\sum_{j=1,j\not=i}^n z_{js}^{(t)}\over\sum_{j=1}^n z_{js}^{(t)}\right)^2+{\sum_{j=1,j\not=i} (z_{js}^{(t)})^2\over(\sum_{j=1}^n z_{js}^{(t)})^2}
\end{equation}
 and 
\begin{equation}
\label{eq:definition of covariance difference}
b_{ss'i}=\left({\sum_{j=1,j\not=i}^n z_{js}^{(t)}\over\sum_{j=1}^n z_{js}^{(t)}}\right)\left({\sum_{j=1,j\not=i}^n z_{js'}^{(t)}\over\sum_{j=1}^n z_{js'}^{(t)}}\right)+{\sum_{j=1,j\not=i} z_{js}^{(t)} z_{js'}^{(t)}\over(\sum_{j=1}^n z_{js}^{(t)})(\sum_{j=1}^n z_{js'}^{(t)})}.
\end{equation}
By the mean and covariance formulae of quadratic terms of normal distributions, there is
\begin{equation}
\label{eq:mean of difference of the distance between two centers in spherical}
\eqalign{
{\rm E}(d_{ss'i})=&p\sigma^2(a_{si}-a_{s'i})+\|{\bm\nu}_{ssi}\|^2+\|{\bm\nu}_{ss'i}\|^2\cr
}\end{equation}
and
\begin{equation}
\label{eq:variance of difference of the distance between two centers in spherical}
\eqalign{
{\rm var}(d_{ss'i})=&2p\sigma^4(a_{si}^2+a_{s'i}^2+2b_{ss'i}^2)+4a_{si}\sigma^2\|{\bm\nu}_{ss'i}\|^2+4a_{s'i}\sigma^2\|{\bm\nu}_{ss'i}\|^2 +8b_{ss'i}\sigma^2\langle{\bm\nu}_{ssi},{\bm\nu}_{ss'i} \rangle.\cr
}
\end{equation}
The first terms on the right-hand sides of~\eqref{eq:mean of difference of the distance between two centers in spherical} and~\eqref{eq:variance of difference of the distance between two centers in spherical} linearly increase with $p$. The remaining terms do not vary with $p$ when $p$ grows but $q$ does not. Thus, the first terms dominate the magnitudes of ${\rm E}(d_{ss'i})$ and ${\rm var}(d_{ss'i})$ if all the variables are used. The difference between mean vectors computed in the M-step can be ignored. The partition given by the $t$th iteration tends to be random if $p\gg q$. A similar issue also occurs in the $k$-means, leading to the following theorems.

\begin{lem}
\label{lem:control the coefficient mean and covariance}
Assume that (i)~\eqref{eq:assumption of spatsity} is satisfied in the spherical GMM, and (ii) either (a) $\sum_{j=1}^n z_{js}^{(t)}\ge 1$ is satisfied for any $s\in\{1,\dots,k\}$ in the iterations of the EM-algorithm or (b)~\eqref{eq:prediction of ground truth latent variables} is used to assign the labels in the $k$-means. Then $\lim_{p\rightarrow\infty}2p\sigma^4(a_{si}^2+a_{s'i}^2+2b_{ss'i}^2)/{\rm var}(d_{ss'i})=1$.
\end{lem}

Lemma~\ref{lem:control the coefficient mean and covariance} assumes that $q$ does not vary with $p$ as $p$ grows. The assumption can be relaxed as that $q$ also grows with a lower order of $p$. It is equivalent to the sparsity assumption adopted by the supervised variable selection problems in the literature. Note that $\sum_{j=1}^n z_{js}^{(t)}$ represents the expected number of observations assigned to cluster $s$ by the $t$th iteration. Condition (ii)(a) requires that the clusters in the iterations of the EM-algorithm contain at least one observation. This condition is always satisfied in the $k$-means, leading to Condition (ii)(b). Thus, only extreme cases are excluded by the assumptions of Lemma~\ref{lem:control the coefficient mean and covariance}. 

\begin{thm}
\label{thm:randomly assignments}
If $\lim_{p\rightarrow\infty} q/p=0$ and Assumptions (i) and (ii) of Lemma~\ref{lem:control the coefficient mean and covariance} are satisfied, then the partition provided by the EM-algorithm or the $k$-means for the spherical GMM tends to be random as $p\rightarrow\infty$.
\end{thm}

We turn our attention to the LDA and QDA GMMs. Because the components of ${\bm x}_i$ may be dependent, we need to address the challenges caused by dependencies. In the literature, asymptotics for dependent data is assessed by functional central limit theorems (FCLTs) with strong mixing conditions. This often involves the strong mixing coefficient between the components of ${\bm x}_i$ defined as $\gamma_l=\sup\{|P(A\cap B)-P(A)P(B)|: A\in\sigma(x_{ij}:1\le j\le m\}, B\in\sigma(x_{ij}: m+l\le j\le p),1\le m\le p-l\}, l\le p-1$. Conditions for $\gamma_l$ are needed. This has been well-studied previously~\cite[e.g.]{herrnodorf1984}. 

\begin{thm}
\label{thm:randomly assignments LDA model}
Assume that (i)~\eqref{eq:assumption of spatsity} is satisfied in the LDA GMM, (ii) $\sum_{j=1}^n z_{js}^{(t)}\ge 1$ for any $s\in\{1,\dots,k\}$ in the iterations of the EM-algorithm, (iii) there exists $c_1,c_2>0$ such that the eigenvalues of ${\bm\Sigma}$ is between $c_1$ and $c_2$, and (iv) $\sup_{1\le l\le p}l^2|\gamma_l|$ is uniformly bounded as $p$ grows. Then the partition given by the EM-algorithm for the LDA GMM tends to be random as $p\rightarrow\infty$.
\end{thm}

\begin{thm}
\label{thm:randomly assignments QDA model}
If all assumption of Theorem~\ref{thm:randomly assignments LDA model} are satisfied, then the partition provided by the EM-algorithm for the QDA GMM tends to be random as $p\rightarrow\infty$.
\end{thm}

\begin{cor}
\label{cor:compared with random clustering unsupervised variable selection}
If $\alpha={\mathcal A}$ and $\lambda=0$ is used in the FPL specified by~\eqref{eq:penalized likelihood loss spherical} for the spherical GMM,~\eqref{eq:penalized likelihood loss lda} for the LDA GMM, or a similar FPL for the QDA GMM, then the cluster assignments provided by the EM-algorithm or the $k$-means tends to be random as $p\rightarrow\infty$.
\end{cor}

Corollary~\ref{cor:compared with random clustering unsupervised variable selection} means that using the likelihood function to select the variables is inappropriate. Therefore, a penalized likelihood approach is needed. Because the usage of the entire variable set is close to be a random assignment of the clusters, it is important to carry out an unsupervised variable selection method to select an active variable set such that reasonable cluster assignments can be induced.  

\section{Simulation}
\label{sec:simulation}

We compare our method with our competitors via Monte Carlo simulations. We evaluate two issues. The first is our finding that clustering can be significantly worse if many uninformative variables are used. The second is our statement that our method selects active variables but our competitors select the union of active and redundant variables, implying that our method can distinguish the active, redundant, and uninformative variables, but our competitors cannot. 

We investigated the first issue based on a simulation with $1000$ replications. We simulated $1000$ datasets with ${\bm x}_i\in\mathbb{R}^{1000}$, $k=2$, and $q=2$, implying that the GMM had two clusters with two relevant and $998$ uninformative variables. We independently generated $n_s=10$ observations from ${\cal N}({\bm\mu}_s,{\bf I})$ for $s=1,2$. We chose $\mu_{11}=\mu_{12}=-\mu_{21}=-\mu_{22}=1.6$ and $\mu_{sj}=0$ for all $j\ge 3$ in the simulation, indicating that the GMM satisfied $\|{\bm\mu}_1-{\bm\mu}_2\|^2=20.48$. If all the variables were considered, then we had ${\rm E}(Q)=22.48$ and ${\rm var}(Q)=73.62$ in~\eqref{eq:expected value for next assignment k-means} and~\eqref{eq:variance for next assignment k-means}. The contributions of the second terms for the difference between ${\bm\mu}_1$ and ${\bm\mu}_2$ in the expressions of ${\rm E}(Q)$ and ${\rm var}(Q)$ were less than $10\%$ and $1\%$, respectively. If only the first two variables were considered, then the contributions increased to $98.1\%$ and $84.9\%$, respectively. 

\begin{figure}
\centerline{\rotatebox{270}{\psfig{figure=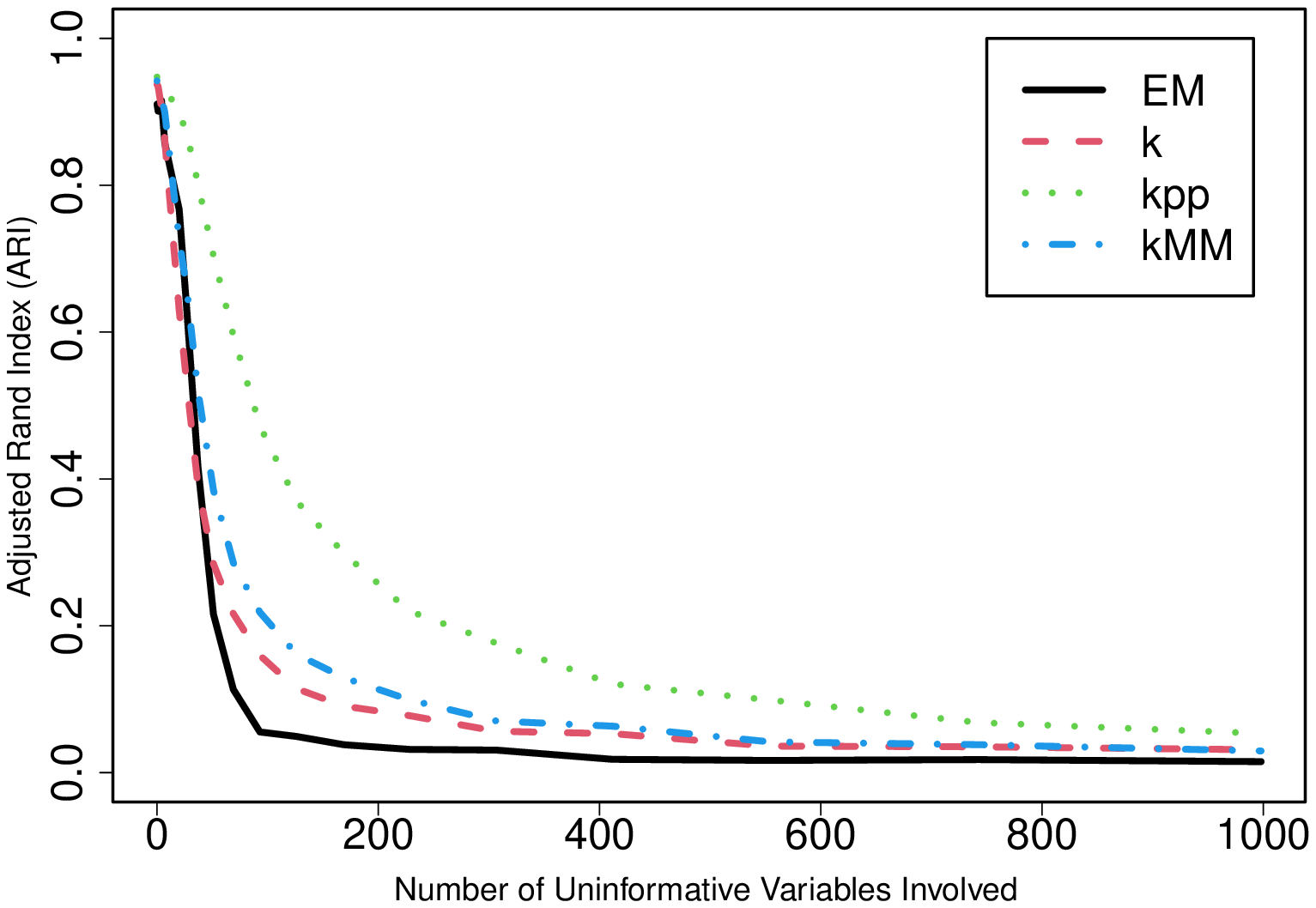,width=3.0in,}}}
\caption{\label{fig:ARI values as number of variables change} Mean ARI values as functions of the number of uninformative variables involved in the EM-algorithm (EM) for spherical GMMs, the $k$-means (k), the $k$-means++ (kpp), and the $k$-meansMM (kMM) derived based on simulations with $1000$ replications.}
\end{figure}

We used the EM-algorithm and the $k$-means to partition the data. We studied three variants of the $k$-means. They were the basic $k$-means, the $k$-means++~\cite{arthur2007}, and the $k$-meansMM~\cite{huang2023,zhang2022}. The iterations of the three variants are the same. The only difference is the initialized centroids. The basic $k$-means uses the random initialization. The $k$-means++  chooses its $k$ initial centroids sequentially. It selects the next initial centroid based on a probability distribution inversely proportional to the minimum squared distance to the previous initial centroids. The $k$-meansMM also chooses its $k$ initial centroids sequentially. It selects the next centroid by maximizing the minimum distance to the previous initial centroids. We applied those methods based on a variable set composed of the first $r$ variables with $r$ varying from $2$ to $1000$. We used the adjusted rand index (ARI) to evaluate the precision of clustering. We found that the ARI values obtained from all of the four methods decreased as the number of uninformative variables increased (Figure~\ref{fig:ARI values as number of variables change}). This confirmed our findings in Sections~\ref{subsec:motivated example} and~\ref{subsec:theoretical evaluation}.

\begin{table}
\scriptsize
\caption{\label{tab:ARI and SSE simulation} Average variable size and the ARI and MSE values provided by the proposed method based on the EM-algorithm, the basic $k$-means (k), the $k$-means++ (kpp), and the $k$-meansMM (kMM) with the comparison to the previous \textsf{sparcl}, \textsf{clustvarsel}, and \textsf{VarSelLCM} methods derived based on simulations with $1000$ replications, where the number of relevant variables is $4$ if $\phi=0.0$ or $8$ if $\phi>0.0$.}
\begin{center}
\begin{tabular}{ccccccccccc}\hline
True  &  Clustering & \multicolumn{3}{c}{$\phi=0.0$}  & \multicolumn{3}{c}{$\phi=1.0$}   & \multicolumn{3}{c}{$\phi=2.0$} \\\cline{3-11}
Model &  Method  & Size & ARI & MSE &  Size & ARI & MSE & Size & ARI & MSE \\\hline
Spherical  & \textsf{GMM}  & $2.80$&$0.9986$&$0.9605$&$2.82$&$0.9984$&$0.9616$&$2.75$&$0.9994$&$0.9600$  \\
   & \textsf{k} &  $3.86$&$0.9836$&$0.9796$&$4.13$&$0.9788$&$0.9878$&$3.86$&$0.9781$&$0.9915$ \\
 &\textsf{kpp}  & $3.79$&$0.9843$&$0.9718$&$3.49$&$0.9914$&$0.9670$&$3.57$&$0.9854$&$0.9723$ \\
 & \textsf{kMM} & $2.76$&$0.9977$&$0.9605$&$2.72$&$0.9982$&$0.9617$&$2.69$&$0.9993$&$0.9600$  \\
 &\textsf{sparcl}&  $4.00$ & $0.9432$ &$1.0042$ & $7.94$ & $0.9264$ & $1.0056$ & $8.00$ & $0.9296$ & $1.0109$ \\
&  \textsf{clustvarsel}& $4.00$ & $0.9990$& $0.9586$ & $7.77$ & $0.9976$ &  $09695$ & $8.00$ & $0.9998$ & $ 0.9607$  \\
& \textsf{VarSelLCM} &  $4.00$ & $0.9392$ & $1.0291$ & $7.74$ & $0.9694$ & $1.0437$ & $7.99$ & $0.9856$ & $1.0404$\\\hline
LDA  & \textsf{GMM}  & $2.82$&$0.9982$&$0.9580$&$2.78$&$0.9974$&$0.9640$&$2.86$&$0.9979$&$0.9634$  \\
    &\textsf{k} &  $3.89$&$0.9814$&$0.9798$&$3.85$&$0.9745$&$0.9907$&$3.93$&$0.9852$&$0.9848$\\
 &\textsf{kpp}  & $3.59$&$0.9879$&$0.9658$&$3.60$&$0.9871$&$0.9720$&$3.91$&$0.9925$&$0.9686$ \\
 & \textsf{kMM} & $2.71$&$0.9976$&$0.9584$&$2.78$&$0.9971$&$0.9640$&$2.81$&$0.9981$&$0.9634$  \\
& \textsf{sparcl}&  $4.00$ & $0.9407$ & $1.0075$  & $7.91$  & $0.9374$ & $0.9867$ & $8.00$ & $0.9293$ & $1.0226$ \\
&  \textsf{clustvarsel}&  $4.02$ & $0.9968$ &$0.9618$  & $7.62$ & $0.9972$ & $0.9478$ & $7.97$ & $0.9994$ &  $0.9630$ \\
& \textsf{VarSelLCM} &  $3.96$ & $0.9317$ & $1.0441$ & $7.60$ & $0.9598$ &$0.9957$ & $7.97$ & $0.9739$ & $1.0165$ \\\hline
\end{tabular}
\end{center}
\end{table}
\normalsize

We then investigated the second issue. It was also based on simulations with $1000$ replications. We simulated $1000$ datasets with ${\bm x}_i\in\mathbb{R}^{50}$, $k=10$, and all $n_s=25$. We had $n_s=25$ for $s=1,\dots,10$. The total number of observations was $n=250$. The total number of variables was $p=50$. To obtain the $10$ clusters, we generated the first and next four components of ${\bm\mu}_s$ independently from ${\cal N}(0,10)$ and ${\cal N}(0,\phi^2)$, respectively. We set the remaining components of ${\bm\mu}_s$ to be $0$. We considered two scenarios after ${\bm\mu}_1,\cdots,{\bm\mu}_{10}$ were obtained. In the first scenario, we independently generated $25$ observations from ${\cal N}({\bm\mu}_s,{\bf I})$. The true distribution of the $s$th cluster was ${\bm x}_i|z_{ri}=1\sim^{iid}{\cal N}({\bm\mu}_s,{\bf I})$. The true model was the spherical GMM. In the second scenario, we generated data by a three-step procedure. In the first step, we generated $50$ eigenvalues independently from ${\cal U}(0,2)$. In the second step, we generated a random orthogonal matrix. We treated the columns of the orthogonal matrix as the eigenvectors. We use the eigenvectors and the eigenvalues to compute the random variance-covariance matrix ${\bm\Sigma}$. In the third step, we independently generated $25$ observations from ${\cal N}({\bm\mu}_s,{\bm\Sigma})$ for every $s=1,\dots,10$. The true distribution of the $s$th cluster was ${\bm x}_i|z_{ri}=1\sim^{iid}{\cal N}({\bm\mu}_s,{\bm\Sigma})$. The true model was the LDA GMM. 

We applied our method and our competitors. In our method, we computed $\hat\alpha$ by~\eqref{eq:selection of variable set} using the EM-algorothm, the basic $k$-means, the $k$-means++, and the $k$-meansMM methods, respectively. To compare, we also considered the previous sparse clustering (\textsf{sparcl})~\cite{witten2010}, the first model-based variable selection clustering (\textsf{clustvarsel})~\cite{raftery2006}, the second model-based variable selection clustering \textsf{VarSelLCM}~\cite{marbac2017}, and the variable selection for clustering and classification (\textsf{vscc})~\cite{andrews2014} methods. The application of the \textsf{vscc} method failed because it could not be used when $k\ge 10$. 

We computed the average size of selected variable sets and the average ARI and MSE values of the resulting partitions by simulations with $1000$ replications (Table~\ref{tab:ARI and SSE simulation}). The results showed that the performance of the $k$-meansMM was better than the other two variants of the $k$-means. It was equally good as the EM-algorithm. For $|\hat\alpha|$, the four variants of our method selected $2$, $3$, or $4$ variables. The number of relevant variables was $4$ when $\phi=0$ or $8$ when $\phi>0$, implying that the number of active variables was lower than the number of relevant variables. It was unnecessary to use all the relevant variables to partition the data. The previous methods selected all the relevant variables. They did not distinguish the active and redundant variables. 

\section{Application}
\label{sec:application}

We applied our method with the comparison to our competitors to a gene expression dataset for cancer RNA sequencing. The dataset can be freely accessed via the UCI Machine Learning Repository website. It contained $n=801$ observations and $p=16,\!383$ variables. The variables are denoted as ${\bm x}_1,\dots,{\bm x}_{p}$, respectively. The observations were derived by a random extraction of gene expression of cancer patients having five different types of tumors denoted as \textsf{BRCA}, \textsf{KIRC}, \textsf{COAD}, \textsf{LUAD}, and \textsf{PRAD}. We treated those as the ground truths and used them to compute the ARI. The corresponding tumor counts were $136$, $141$, $300$, $146$, and $78$, respectively. The goal was to determine the types of tumors by the variables. Both supervised and unsupervised ML approaches can be used to analyze the data. We chose the unsupervised ML approach. 

The number of variables is large. It is important to carry out an unsupervised variable selection procedure to select variables. We applied our proposed and the previous \textsf{sparcl}, \textsf{clustvarsel}, and \textsf{VarSelLCM} methods. We studied the case when $k=5$ and also considered the case when $k$ was determined by the Gap Statistic~\cite{tibshirani2001}. We found that we could use the Gap Statistics to determine the optimal $k$ with $k_{best}=5$ in our method. It might not work well if our competitors were used. We then decided to report the result based on $k=5$.  

Our method with the EM-algorithm selected $9$ variables with $ARI=0.989$. The $9$ variables were ${\bm x}_{400}$, ${\bm x}_{2748}$, ${\bm x}_{4275}$, ${\bm x}_{6817}$, ${\bm x}_{10291}$, ${\bm x}_{11125}$, ${\bm x}_{11465}$, ${\bm x}_{11551}$, and ${\bm x}_{15901}$. Our method with the basic $k$-means selected $10$ variables with $ARI=0.987$. The $10$ variables were ${\bm x}_{198}$, ${\bm x}_{5408}$, ${\bm x}_{7965}$, ${\bm x}_{8014}$, ${\bm x}_{8832}$, ${\bm x}_{9565}$, ${\bm x}_{10280}$, ${\bm x}_{11260}$, ${\bm x}_{11394}$, and ${\bm x}_{15901}$. Our method with the $k$-means++ selected $8$ variables with $ARI=0.976$. The $8$ variables were ${\bm x}_{224}$, ${\bm x}_{1190}$, ${\bm x}_{4179}$, ${\bm x}_{7965}$, ${\bm x}_{9178}$, ${\bm x}_{11551}$, ${\bm x}_{11911}$, and ${\bm x}_{15899}$. Our method with the kMM selected $8$ variables with $ARI=0.994$. The $8$ variables were 
 ${\bm x}_{1190}$, ${\bm x}_{4179}$, ${\bm x}_{6877}$, ${\bm x}_{7114}$, ${\bm x}_{7624}$, ${\bm x}_{7965}$, ${\bm x}_{8004}$, and ${\bm x}_{8108}$. The selected variable sets were different from the four variants of our method because of the non-uniqueness issue of the active and redundant variables in the unsupervised variable selection problem. 

We then considered our competitors. The \textsf{sparcl} method selected $63$ variables with $ARI=0.980$. The \textsf{sparcl} method did not select variables directly. It provided a weight for each variable with a larger weight indicating that the variable was more important. We indirectly selected the variables if the corresponding weights were higher than $0.001$. The \textsf{clustvarsel} method selected $17$ variables with $ARI=0.037$. We investigated this issue and found that the \textsf{clustvarsel} method preferred variables with only a few distinct values. For instance, it selected ${\bm x }_{10}$ because ${\bm x}_{10}$ only had five different options. Similar issues also occurred in the \textsf{vscc} method. This phenomenon did not occur in our proposed and the previous \textsf{sparcl} method. The \textsf{VarSelLCM} method failed in the application. Based on the comparison, we concluded that our method outperformed the previous methods. 

\section{Discussion}
\label{sec:discussion}

We expect that our unsupervised variable selection method can be implemented in arbitrary clustering methods beyond the GMM and the $k$-means. The performance of a clustering method based on a subset of variables can be better than that based on the set of all variables. Variable selection can enhance the clustering accuracy if only a subset of relevant variables is used. Although we only investigate the GMM and the $k$-means, we believe that our findings are also true if other clustering methods are used. This is left to future research.


\appendix
\section{Proofs}

\noindent
{\bf Proof of Theorem~\ref{thm:uninformative GMM}.} Suppose that ${\bm\mu}_s={\rm E}({\bm x}_i|z_{is}=1)$ can be expressed as ${\bm\mu}_s=({\bm\mu}_{su}^\top,{\bm\mu}_{sr}^\top)^\top$ such that it satisfies ${\rm E}({\bm x}_{iu}|z_s=1)={\bm \mu}_{su}$, ${\rm E}({\bm x}_{ir}|z_{is}=1)={\bm \mu}_{sr}$. In addition, suppose that ${\bm\Sigma}_s={\rm cov}({\bm x}_i|z_{is}=1)$ can be expressed as
$${\bm\Sigma}_s=\left( \begin{array}{cc} {\bm\Sigma}_{suu} & {\bm\Sigma}_{sur} \cr {\bm\Sigma}_{sru} & {\bm\Sigma}_{srr}  \end{array}  \right),$$
where ${\rm var}({\bm x}_{iu}|z_{is}=1)={\bm\Sigma}_{ruu}$, ${\rm cov}({\bm x}_{iu},{\bm x}_{ir}|z_{is}=1)={\bm\Sigma}_{sur}={\bm\Sigma}_{sru}^\top$, and ${\rm var}({\bm x}_{ir}|z_{is}=1)={\bm\Sigma}_{srr}$. We obtain ${\bm x}_{iu}|{\bm x}_{ir},z_{is}=1\sim {\cal N}[{\bm\mu}_{su}- {\bm\Sigma}_{sur}{\bm\Sigma}_{srr}^{-1} ({\bm x}_{ir}-{\bm\mu}_r),  {\bm\Sigma}_{suu}-{\bm\Sigma}_{sur}{\bm\Sigma}_{srr}^{-1}{\bm\Sigma}_{sur}^\top]$, leading to the first conclusion of the theorem. The first conclusion implies that the union of two distinct uninformative set of variables is uninformative. We search for the largest ${\mathcal A}_u$ to satisfy the corresponding conditions, leading to the second conclusion. \qed

\noindent
{\bf Proof of Lemma~\ref{lem:control the coefficient mean and covariance}.} Let $u_i=z_{is}^{(t)}/\sum_{j=1}^n z_{js}^{(t)}$ and $v_i=z_{is'}^{(t)}/\sum_{j=1}^n z_{js'}^{(t)}$. Then, $a_{si}=(1-u_i)^2+\sum_{j=1,j\not=i}^n u_j^2$ and $a_{s'i}=(1-v_i)^2+\sum_{j=1,j\not=i}^n v_j^2$. By $z_{is}^{(t)}+z_{is'}^{(t)}\le 1$, we obtain $u_i+v_i\le 1$. Thus, $a_{si}^2+a_{s'i}^2\ge (1-u_i)^4+(1-v_i)^4\ge (1-u_i)^4+u_i^4\ge \min_{u\in[0,1]}\{(1-u)^4+u^4\}=1/8$, implying that the first term on the right-hand side of~\eqref{eq:variance of difference of the distance between two centers in spherical} linearly increases with $p$. Because $\sum_{j=1}^n u_j=\sum_{j=1}^n v_j=1$, $a_{si}$, $a_{s'i}$, and $b_{ss'i}$ are all less than $2$. Because the $j$th components of ${\bm\nu}_{ssi}$ and ${\bm\nu}_{ssi'}$ are zero if $j\ge q$, the summation of the second, third, and fourth term on the right-hand side of~\eqref{eq:variance of difference of the distance between two centers in spherical} is uniformly bounded as $p\rightarrow\infty$. We conclude. \qed

\noindent
{\bf Proof of Theorem~\ref{thm:randomly assignments}.} As the components of each ${\bm x}_i$ can be treated as random samples from independent normal distributions with different means, 
 asymptotic normality of quadratic forms of normal distributions can be applied. We conclude by combining it with Lemma~\ref{lem:control the coefficient mean and covariance}. \qed

\noindent
{\bf Proof of Theorem~\ref{thm:randomly assignments LDA model}.} To simplify our notations, we use $\hat{\bm\Sigma}$ to represent ${\bm\Sigma}^{(t)}$ for the $t$th iteration of the EM-algorithm for the LDA GMM. After a few steps of algebra, we find that the label assignments given by $t$th iteration are determined by $\tilde d_{ss'i}={\bm r}_{si}^\top\hat{\bm\Sigma}^{\ddagger}{\bm r}_{si}-{\bm r}_{s'i}^\top\hat{\bm\Sigma}^{\ddagger}{\bm r}_{s'i}$ for distinct $s,s'\in\{1,\dots,k\}$, where ${\bf A}^\ddagger$ represents the Moore-Penrose generalized inverse of matrix ${\bf A}$. Let ${\bm u}_{si}={\bm\Sigma}^{-1/2}{\bm r}_{si}$. Then, we have $\tilde d_{ss'i}={\bm u}_{si}^\top({\bm\Sigma}^{-1/2}\hat{\bm\Sigma}{\bm\Sigma}^{-1/2})^{\ddagger}{\bm u}_{si}-{\bm u}_{s'i}^\top({\bm\Sigma}^{-1/2}\hat{\bm\Sigma}{\bm\Sigma}^{-1/2})^{\ddagger}{\bm u}_{s'i}$ and ${\bm\Sigma}^{-1/2}\hat{\bm\Sigma}{\bm\Sigma}^{-1/2}=\sum_{i=1}^n {\bm u}_{si}{\bm u}_{si}^\top$. Using the mean and covariance formulations for normal distributions, we have ${\rm E}({\bm u}_{si})={\bm\Sigma}^{-1/2}{\bm\nu}_{ssi}$, ${\rm E}({\bm u}_{s'i})={\bm\Sigma}^{-1/2}{\bm\nu}_{ss'i}$, ${\rm var}({\bm u}_{si})=a_{si}{\bf I}$, ${\rm var}({\bm u}_{s'i})=a_{s'i}{\bf I}$, and ${\rm cov}({\bm u}_{si},{\bm u}_{s'i})=b_{ss'i}{\bf I}$.  For each $i$, we can treat $x_{i1},\dots,x_{ip}$ as a time series as $p$ growth. Because (iv) is satisfied, we can use the FCLT for weakly dependent data~\cite[Corollay 2]{herrnodorf1984} to show our conclusion. By the continuous mapping theorem, we conclude that the asymptotic distribution of $\tilde d_{ss'i}/p$ is identical to that of $(\|{\bm u}_{si}\|^2-\|{\bm u}_{s'i}\|^2)/p$ as $p\rightarrow\infty$.  By (i) and (iii), we have $\|{\bm\Sigma}^{-1/2}{\bm\nu}_{ssi}\|^2\le c_1^{-1}\|{\bm\nu}_{ssi}\|^2=O(q)$ and $\|{\bm\Sigma}^{-1/2}{\bm\nu}_{ss'i}\|^2\le c_1^{-1}\|{\bm\nu}_{ss'i}\|^2=O(q)$, implying that the difference between ${\bm\mu}_s$, $s=1,\dots,k$, can be ignored. We conclude. \qed

\noindent
{\bf Proof of Theorem~\ref{thm:randomly assignments QDA model}.} We can directly use the proof of Theorem~\ref{thm:randomly assignments LDA model}. The detail is omitted. \qed

\noindent
{\bf Proof of Corollary~\ref{cor:compared with random clustering unsupervised variable selection}.} This is obvious from the theorems. \qed


\begin{thebibliography}{}

\bibitem{andrews2014}
Andrews, J. and McNicholas, P.D. (2014). Variable selection for clustering and classification. {\it Journal of Classification}, {\bf 31}, 136-153.
\bibitem{arias-castro2017}
Arias-Castro, E., and Pu, X. (2017). A simple approach to sparse clustering. {\it Computational Statistics and Data Analysis}, {\bf 105}, 217–228.
\bibitem{arthur2007}
Arthur, D. and Vassilvitskii, S. (2007). k-means++: the advantages of careful seeding. Proceedings of the {\it 18th annual ACM-SIAM symposium on Discrete algorithms}. Society for Industrial and Applied Mathematics Philadelphia, PA, USA. pp. 1027–1035.
\bibitem{bock2008}
Bock, H. (2008). Origins and extensions of the k-means algorithm in cluster analysis electron. {\it Electronic Journal for History of Probability and Statistics}, {\bf 4}, Article 14.
\bibitem{brusco2001}
Brusco, M. and Cradit, J.D. (2001). A variable-selection heuristic for K-means clustering. {\it Psychometrika}, {\bf 66}, 249-270.
\bibitem{cardot2012}
Cardot, H., C\'enac, P., and Monnez, J.M. (2012). A fast and recursive algorithm for clustering large datasets with $k$-medians. {\it Computational Statistics and Data Analysis}, {\bf 6}, 1434-1449.
\bibitem{chan2004}
Chan, E.Y., Ching, W.K., Ng, M.K., and Huang, J.Z. (2004). An optimization algorithm for clustering using weighted dissimilarity measures. {\it Pattern recognition}, {\bf 37}, 943-952.
\bibitem{chaturvedi2001}
Chaturvedi, A., Green, P.E., and Caroll, J.D., (2001). K-modes clustering. {\it Journal of Classification}, {\bf 18}, 35-55.
\bibitem{cheung2006}
Cheung, Y., and Zeng, H. (2006) Feature weighted rival penalized em for Gaussian mixture clustering: automatic feature and model selections in a single paradigm. IEEE 2006 International Conference on Computational Intelligence and Security, {\bf 1}, pp. 633-638.
\bibitem{dempster1977}
Dempster, A., Laird, N., and Rubin, D. (1997). Maximum likelihood from incomplete data via the EM algorithm (with discussion). {\it Journal of Royal Statistical Society Series B}, {\bf 39}, 1-38.
\bibitem{du2002}
Du, Q., and Wong, T.W. (2002). Numerical studies for MacQueen's $k$-means algorithms for computing the centroidal Voronoi tessellations. {\it Computer and Mathematics with Applications}, {\bf 44}, 511-523.
\bibitem{ester1996}
Ester, M., Kriegel, H.P., Sander, J., and Xu, X (1996). A density-based algorithm for discovering clusters in large spatial databases with noise. In kdd, {\bf 96}, 226-231.
\bibitem{fanli2001}
Fan, J. and Li, R. (2001). Variable selection via nonconcave penalized likelihood and its oracle properties. {\it Journal of the American Statistical Association}, {\bf 96}, 1348-1360.
\bibitem{fop2018}
Fop, M. and Murphy, T.B. (2018). Variable selection methods for model-based clustering. {\it Statistics Surveys}, {\bf 12}, 18-65.
\bibitem{fowlkes1988}
Fowlkes, E. B., Gnanadesikan, R., and Kettering, J. R. (1988). Variable selection in clustering. {\it Journal of Classification}, {\bf 5}, 205–228.
\bibitem{friedman2004}
Friedman, J.H., and Meulman, J.J. (2004). Clustering objects on subsets of attributes (with discussion). {\it Journal of the Royal Statistical Society Series B}, {\bf 66}, 815–849.
\bibitem{fraley2002}
Fraley, C. and Raftery, A.E. (2002). Model-based clustering, discriminant analysis, and density estimation. {\it Journal of the American Statistical Association}, {\bf 97}, 611-631.
\bibitem{gnanadesikan1995}
Gnanadesikan, R., Kettenring, J. R., and Tsao, S. L. (1995). Weighting and selection of variables for cluster analysis. {\it Journal of Classification}, {\bf 12}, 113-136.
\bibitem{hancer2020}
Hancer, E., Xue, B., and Zhang, M. (2020). A survey on feature selection approaches for clustering. {\it Artificial Intelligence Review}, {\it 53}, 4519–4545.
\bibitem{herrnodorf1984}
Herrnodorf, H. (1984). A functional central limit theorem for weakly dependent sequences of random variables. {\it Annals of Probability}, {\bf 12}, 141-153
\bibitem{huang2005}
Huang, J.Z., Ng, M.K., Rong, H, and Li, Z.(2005). Automated variable weighting in k-means type clustering. {\it IEEE transactions on pattern analysis and machine intelligence}, {\bf 5}, 657-668.
\bibitem{huang2023}
Huang, H., Tang, Z., Zhang, T., and Yang, B. (2023). Improved clustering using nice initialization. In GLOBECOM 2023-2023 IEEE Global Communications Conference (pp. 320-325).
\bibitem{jing2007}
Jing, L., Ng, M.K., and Huang, J. (2007). An entropy weighting k-means algorithm for subspace clustering of high-dimensional sparse data. {\it IEEE Transactions on knowledge and data engineering}, {\bf 19}, 1026-1041.
\bibitem{kriegel2011}
Kriegel, H.P., Kr\"oger, P., Sander, J., and Zimek, A. (2011). Density-based clustering. {\it WIRES Data Mining Knowledge Discovery}, {\bf 1}, 231-240.
\bibitem{lau2007}
Lau, J.W. and Green, P.J. (2007). Bayesian model-based clustering procedures. {\it Journal of Computational and Graphical Statistics}, {\bf 16}, 526-558.
\bibitem{li2017}
Li, J., Cheng, K., Wang, S., Morstatter, F., Trevino, R.P., Tang, J., and Liu, H. (2017). Feature selection: A data perspective. {\it ACM computing surveys (CSUR)}, {\bf 50}, 1-45.
\bibitem{macqueen1967}
MacQueen, J. B. (1967). Some methods for classification and analysis of multivariate observations. {\it Proceedings of 5th Berkeley Symposium on Mathematical Statistics and Probability}. University of California Press. pp. 281–297.
\bibitem{mahajan2012}
Mahajan, M., Nimbhorkar, P., and Varadarajan, K. (2012). The planar k-means problem is NP-hard. {\it Theoretical Computer Science}, {\bf 442}, 13-21.
\bibitem{marbac2017}
Marbac, M., and Sedki, M. (2017). Variable selection for model-based clustering using the integrated complete-data likelihood. {\it Statistics and Computing}, {\bf 27}, 1049-1063.
\bibitem{maugis2009a}
Maugis, C., Celeus, G., and Martin-Magniette, M.L. (2009a). Variable selection for clustering with Gaussian mixture models. {\it Biometrics}, {\bf 65}, 701-709.
\bibitem{maugis2009b}
Maugis, C., Celeus, G., and Martin-Magniette, M.L. (2009b). Variable selection in model-based clustering: a general variable role modeling. {\it Computational Statistics and Data Analysis}, {\bf 53}, 3872-3882.
\bibitem{milligan1989}
Milligan, G. W. (1989). A validation study of a variable weighting algorithm for cluster analysis. {\it Journal of Classification}, {\bf 6}, 53–71.
\bibitem{nityasuddhi2003}
Nityasuddhi, D., and B\"ohning, D. (2003). Asymptotic properties of the EM algorithm estimate for normal mixture models with component specific variances. {\it Computational Statistics and Data Analysis}, {\bf 41}, 591-601.
\bibitem{parvin2015}
Parvin, H., and Minaei-Bidgoli, B. (2015). A clustering ensemble framework based on selection of fuzzy weighted clusters in a locally adaptive clustering algorithm. {\it Pattern Analysis and Applications}, {\bf 18}, 87-112.
\bibitem{raftery2006}
Raftery, A.E, and Dean, N. (2006). Variable selection for model-based clustering. {\it Journal of the American Statistical Association}, {\bf 101}, 168-178.
\bibitem{steinley2008a}
Steinley, D., and Brusco, M. J. (2008a). A new variable weighting and selection procedure for k-means cluster analysis. {\it Multivariate Behavioral Research}, {\bf 43}, 77–108.
\bibitem{steinley2008b}
Steinley, D., and Brusco, M. J. (2008b). Selection of variables in cluster analysis: An empirical comparison of eight procedures. {\it Psychometrika}, {\bf 73}, 125.
\bibitem{steinhaus1957}
Steinhaus, H. (1957). Sur la division des corps matériels en parties. {\it Bulletin L’Académie Polonaise des Science}, {\bf 4}, 801-804, (In French).
\bibitem{tang2022}
Tang, Z., Zhang, T., Yang, B., Su, J., and Song, Q. (2022). SiGra: Single-cell spatial elucidation through image-augmented graph transformer. bioRxiv. doi: https://doi.org/10.1101/2022.08.18.504464
\bibitem{tibshirani1996}
Tibshirani, R.J. (1996). Regression shrinkage and selection via the LASSO. {\it Journal of the Royal Statistical Society Series B}, {\bf 58}, 267-288.
\bibitem{tibshirani2001}
Tibshirani, R., Walther, G., and Hastie (2001). Estimating he number of clusters in hart via the gap statistic. {\it Journal of Royal Statistical Society Series B}, {\bf 63}, 411-423.
\bibitem{trauwaert1991}
Trauwaert, E., Kaufman, L., and Rousseeuw, P. (1991). Fuzzy clustering algorithms based on the maximum likelihood principle. {\it Fuzzy Sets and Systems}, {\bf 42}, 213-227.
\bibitem{witten2010}
Witten, D.M., and Tibshirani, R. (2010). A framework for feature selection in clustering. {\it Journal of the American Statistical Association}, {\bf 105}, 713-726.
\bibitem{yuan2022}
Yuan, S., De Roover, K., and van Deun, K. (2022). Simultaneous clustering and variable selection: A novel algorithm and model selection procedure. {\it Behavior Research Methods}, 1-18, Doi: 10.3758/s13428-022-01795-7.
\bibitem{zhang2010}
Zhang, C.H. (2010). Nearly unbiased variable selection under minimax concave penalty. {\it Annals of Statistics}, {\bf 38}, 894-942.
\bibitem{zhangli2010}
Zhang, Y., Li, R., and Tsai, C. (2010). Regularization parameter selections via generalized information criterion. {\it Journal of the American Statistical Association}, {\bf 105}, 312-323.
\bibitem{zhanglin2021}
Zhang, T., and Lin, G. (2021). Generalized $k$-means in GLMs with applications to the outbreak of COVID-19 in the United States. {\it Computational Statistics and Data Analysis}, {\bf 159}, 107217.
\bibitem{zhang2022}
Zhang, T. (2022). Asymptotics for the $k$-means. arXiv preprint arXiv:2211.10015.
\bibitem{zhao2005}
Zhao, Y. and Karypis, G. (2005). Hierarchical clustering algorithms for document datasets. {\it Data Mining and Knowledge Discovery}, {\bf 10}, 141-168.

\end{thebibliography}
\end{document}